\documentclass[aps,pra,twocolumn,tightenlines,showpacs]{revtex4}
\usepackage{amssymb}
\usepackage{epsfig}
\usepackage{amsbsy}
\usepackage{amsmath}
\usepackage{graphicx}

\begin{document}

%\begin{CJK*}{GBK}{song}
\title{Two-photon transport through a waveguide coupling to a whispering gallery resonator containing an atom and photon-blockade effect}
\author{ T. Shi$^{1}$ and Shanhui Fan$^{2}$}
\affiliation{$^{1}$Max-Planck-Institut f\"{u}r Quantenoptik, Hans-Kopfermann-Str. 1, 85748
Garching, Germany  \\
 $^{2}$Ginzton Laboratory, Department of Electrical Engineering, Stanford University, Stanford, California 94305, USA}
\date{\today}

\begin{abstract}
We investigate the two-photon transport through a waveguide side-coupling to a whispering-gallery-atom system. Using the Lehmann-Symanzik-Zimmermann (LSZ) reduction approach, we present the general formula for the two-photon processes including the two-photon scattering matrices, the wavefunctions and the second order correlation functions of the out-going photons. Based on the exact results of the second order correlation functions, we analyze the quantum statistics behaviors of the out-going photons for
two different cases: (a) the ideal case without the inter-modal coupling in the whispering
gallery resonator; (b) the case in the presence of the inter-modal coupling which leads to more complex nonlinear behavior. In the ideal case, we show that the system consists of two independent scattering pathways, a free pathway by a cavity mode without atomic excitation, and a "Jaynes-Cummings" pathway described by the Jaynes-Cummings Hamiltonian of a single-mode cavity coupling to an atom. The free pathway does not contribution to correlated two-photon processes. In the presence of intermodal mixing, the system no longer exhibit a free resonant pathway. Instead, both the single-photon and the two photon transport properties depend on the position of the atom. Thus, in the presence of intermodal mixing one can in fact tune the photon correlation properties by changing the position of the atom. Our formalism can be used to treat resonator and cavity dissipation as well.
\end{abstract}
\maketitle

%\end{CJK*}

\section{Introduction}

\begin{figure}[tbp]
\includegraphics[bb=117 470 504 787, width=6 cm, clip]{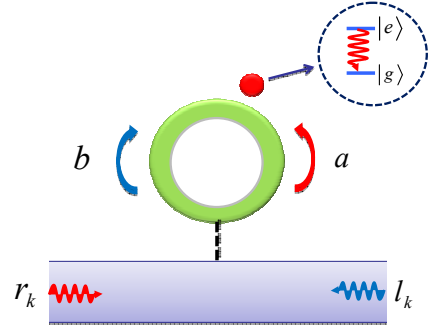}
\caption{(Color online) The schematics for the system: The blue and red
arrows denote the two whispering-gallery modes, which interact with the atom
and the waveguide.}
\label{fig1}
\end{figure}

Recently, the whispering-gallery-atom system inspires a lot of interest \cite%
{WGA1,WGA2,WGA3,WGA4,Theo1,Theo2,Theo3,Theo4,Theo5,Theo6,Theo7}, owing to
its broad applications in the studies of quantum optics. A schematic of such
a system are shown in Fig. \ref{fig1}, where a waveguide is side-coupled to
a whispering-gallery resonator, which then couples to a two-level system. In
the experimental study of this system \cite{WGA2,WGA3,WGA4}, the two-level
system can be either a quantum dot \cite{WGA2}, or an actual atom \cite{WGA3}%
. Here, we will refer such a two-level system as an \textquotedblleft
atom\textquotedblright . These experiments have measured the transmission
properties of such a waveguide-resonator-atom system under weak light
excitation, and have demonstrated quantum effects including anti-bunching
and photon blockade effects.

The experimental progress on this system, in turn, motivated several recent
theoretical studies. Srinivasan and Painter analyzed this system under the
excitation of a weak coherent state input \cite{cl}, and obtained
transmission properties and coherence properties of the transmission through
a numerical procedure with a truncated number-state basis for the photons in
the resonator \cite{Theo1}. Shen and Fan calculated analytically the
transmission properties of such a resonator system with an input of a
single-photon Fock state \cite{Theo2,Theo3}. Subsequently, the single-photon
transports in the whispering-gallery system have been extensively studied
\cite{Theo4,Theo5,Theo6,Theo7}.

In this paper, we study the two-photon transport property of this system
shown in Fig. 1. While the response of the system under weak coherent-state
input yields much information about its non-classical properties, we note
that one important goal of integrated quantum optics is to process quantum
states. Therefore, it is important to study such a system with the input of
non-classical states such as Fock state as well. In addition, in contrast to
the study of the transport properties of a single photon \cite%
{SP1,SP2,SP3,SP4,SP5,SP6}, the studies of two-photon transports provide
important information \cite{MP1,MP2,MP3,MP4,MP5,MP6,MP7,MP8} about
atom-induced photon-photon interaction that is absent in the single-photon
Hilbert space. Of particular interest is the anti-bunching behavior of the
two outgoing photons, which indicate the photon blockade effects \cite{MP6},
as has been observed in a variety of systems \cite%
{AC1,AC2,AC3,AC4,AC5,AC6,PBEF,Kimble,JC2,JC3,SG,TY,PBE,cqed}.

The paper is organized as follows. In Sec. II we review the model
Hamiltonian and our theoretical approach based on the
Lehmann-Symanzik-Zimmermann (LSZ) reduction formula in quantum field theory
\cite{LSZ}. This approach results in an exact formula for the scattering
matrix of the system in multi-photon sub-space. In Sec. III, we present the
exact analytical results, including the single-photon transmissions and the
wavefunctions for the two out-going photons. Based on the general analytical
results, we analyze the very distinct quantum statistics behaviors of the
out-going photons for two different cases in Sec.~IV and Sec.~V. Section IV
considers the ideal case in the absence of inter-modal coupling in the
whispering gallery resonator. Section V analyzes the effect of inter-modal
coupling which leads to more complex nonlinear behavior. In Sec.~VI, the
results are summarized with some remarks and outlooks.

\section{Model Setup}

The system in Fig. 1 is described by the Hamiltonian \cite{Theo2,Theo3}%
\begin{equation}
H=H_{\mathrm{W}}+H_{\mathrm{wg}}+H_{\mathrm{hyb}}.  \label{1}
\end{equation}%
Here,%
\begin{equation}
H_{\mathrm{W}}=\sum_{k>0}kr_{k}^{\dagger }r_{k}-\sum_{k<0}kl_{k}^{\dagger
}l_{k}
\end{equation}%
describes photons propagating in the waveguide. $r_{k}$ ($r_{k}^{\dagger }$)
and $l_{k}$ ($l_{k}^{\dagger }$) are the annihilation (creation) operators
of right-moving and left-moving photons, respectively. The right-moving
photons have momentum $k>0$, and the left-moving photons have momentum $k<0$.

In Eq. (\ref{1})%
\begin{eqnarray}
H_{\mathrm{wg}} &=&\Omega \left\vert e\right\rangle \left\langle
e\right\vert +\omega _{c}(a^{\dagger }a+b^{\dagger }b)+hb^{\dagger
}a+h^{\ast }a^{\dagger }b  \notag \\
&&+\sigma ^{+}(g_{a}a+g_{b}b)+\mathrm{H.c.}  \label{2}
\end{eqnarray}%
describes the two-level system, the resonator, and the interaction between
them under the rotating-wave approximation. In Eq. (\ref{2}), $a$ and $b$
are the annihilation operators for the two counter rotating modes in the
resonator. $h$ is the strength of intermodal-coupling between these two
modes and is typically induced by surface roughness on the resonator. The
atom supports a ground state $\left\vert g\right\rangle $ and an excited
state $\left\vert e\right\rangle $ with a transition frequency $\Omega $. $%
g_{a}$ ($g_{b}$) is the coupling constant between mode $a$ ($b$) and the
atom, and $\sigma ^{+}=\left\vert e\right\rangle \left\langle g\right\vert $.

In Eq. (\ref{1}),%
\begin{equation}
H_{\mathrm{hyb}}=\frac{1}{\sqrt{L}}\sum_{k}(V_{R}r_{k}^{\dagger
}a+V_{L}l_{k}^{\dagger }b+\mathrm{H.c.})
\end{equation}%
describes the coupling between the waveguide and the resonator. Here, we
adopt a box-normalization scheme with $L$ being the length of the waveguide,
and $V_{R}$ ($V_{L}$) is the coupling strength between the right-moving
(left-moving) photon in the waveguide and the mode $a$ ($b$) in the
whispering-gallery system.

We end this section by briefly commenting on various experimental aspects
that are related to the Hamiltonian of Eq. (1). As a representative example,
the experiment in Refs. \cite{WGA2,WGA3,WGA4} has $\omega _{c}\sim \Omega
\sim 10^{6}GHz$. The resonator linewidth $\Gamma /2\pi =0.1\sim 10MHz$. The
atom-resonator coupling constant $\left\vert g_{i}\right\vert /2\pi $ ($%
i=a,b $) can reach $10MHz\sim 10GHz$. While our theory does not depend on
such a detailed choice of parameters, in the numerical examples below we
will be focusing on a similar strong-coupling regime where $\left\vert
g_{i}\right\vert \gg \Gamma $.

In the experiment in Refs. \cite{WGA3} the intermodal mixing has a strength
of $|h|/2\pi =1\sim 100MHz$. The phase of the $h$ depends on the detail of
the surface roughness and typically cannot be controlled in an experiment.
On the other hand, the relative phase of $g_{a}$ and $g_{b}$ depends on the
position of the atom, which can be controlled experimentally \cite%
{Theo2,Theo3}. Since the size of the whispering gallery mode is at least a few
wavelength, the relative phase of $g_{a}$ and $g_{b}$ can vary anywhere
between $0$ and $2\pi $.

\section{Overview of the theoretical approach}

We solve the single- and two-photon scattering matrix ($S$-matrix) for the
Hamiltonian in Eq. (\ref{1}) using the LSZ approach. This approach has been
discussed in details in Ref. \cite{MP5,MP6}. Here we only provide a brief
summary of those aspects that are relevant for subsequent discussions.

The single-photon and two-photon $S$-matrices read%
\begin{equation}
S_{p;k}=\delta _{kp}+iT_{p;k},  \label{S1}
\end{equation}%
and%
\begin{equation}
S_{p_{1}p_{2};k_{1}k_{2}}=S_{p_{1}k_{1}}S_{p_{2}k_{2}}+S_{p_{2}k_{1}}S_{p_{1}k_{2}}+iT_{p_{1}p_{2};k_{1},k_{2}},
\label{S2}
\end{equation}%
respectively, where $k$ ($p$) is the momentum of the incident (out-going)
single-photon, and $k_{1}$, $k_{2}$ ($p_{1}$, $p_{2}$) are the momenta of
two incident (out-going) photons. The right and left moving photons have the
positive and negative momenta, respectively. Here, the $T$-matrix element $%
T_{p_{1},...,p_{n};k_{1},...,k_{n}}\equiv T_{[\mathbf{p;k}]}$ has the form%
\begin{equation}
iT_{[\mathbf{p;k}]}=\lim_{\substack{ \omega _{k_{j}}\rightarrow k_{j},  \\ %
\omega _{p_{j}}\rightarrow p_{j}}}\frac{(2\pi )^{n}G_{[\mathbf{p;k}]}(\omega
_{\mathbf{p}},\omega _{\mathbf{k}})}{\prod\limits_{j=1}^{n}[G_{0}(\omega
_{k_{j}},k_{j})G_{0}(\omega _{p_{j}},p_{j})]}.  \label{T}
\end{equation}%
In Eq. (\ref{T}), $G_{0}(\omega ,k)=i/(\omega -\varepsilon _{k}+i0^{+})$ is
the free propagator of a single photon with $\varepsilon _{k}=\left\vert
k\right\vert $ being the dispersion relation of the waveguide.%
\begin{eqnarray}
G_{[\mathbf{p;k}]}(\omega _{\mathbf{p}},\omega _{\mathbf{k}}) &=&\int
\prod_{j=1}^{n}[\frac{dt_{j}dt_{j}^{\prime }}{2\pi }]G_{[\mathbf{p;k}]}(%
\mathbf{t}^{\prime },\mathbf{t})  \notag \\
&&\times \prod_{j=1}^{n}[\exp (i\omega _{p_{j}}t_{j}^{\prime }-i\omega
_{k_{j}}t_{j})]
\end{eqnarray}%
is the Fourier transform of the exact Green function $G_{[\mathbf{p;k}]}(%
\mathbf{t}^{\prime },\mathbf{t})$. Such a Green function can be determined
as:
\begin{eqnarray}
&&G_{[\mathbf{p;k}]}(\mathbf{t}^{\prime },\mathbf{t})  \label{G} \\
&=&\left. \frac{(-)^{n}\delta ^{2n}\ln Z[\eta _{k},\eta _{k}^{\ast }]}{%
\delta \eta _{f,p_{1}}^{\ast }(t_{1}^{\prime })\text{...}\delta \eta
_{f,p_{n}}^{\ast }(t_{n}^{\prime })\delta \eta _{f,k_{1}}(t_{1})\text{...}%
\delta \eta _{f,k_{n}}(t_{n})}\right\vert _{\substack{ \eta _{f,k}=0  \\ %
\eta _{f,k}^{\ast }=0}},  \notag
\end{eqnarray}%
where
\begin{equation}
Z[\eta _{k},\eta _{k}^{\ast }]=\int D[\mathrm{field}]\exp (iS_{\mathrm{c}%
}+iS_{\mathrm{ex}})  \label{Z}
\end{equation}%
is the generating functional in the path integral formalism. Here, $\int D[%
\mathrm{field}]$ denotes the path integral over the all fields in the
system, $S_{\mathrm{c}}$ is the action of the system, and%
\begin{equation}
S_{\mathrm{ex}}=\sum_{f=r,l}\int dt\sum_{k}[\eta _{f,k}^{\ast }(t)f_{k}(t)+%
\mathrm{H.c.}].
\end{equation}%
describes the external sources $\eta _{f,k}(t)$ that injects photons into
the waveguide.

Based on the formalism as outlined above, in the next section we show the
exact analytical results for the single-photon and two-photon scattering
processes.

\section{Analytic Results for One and Two Photon $S$-matrices}

In this section, the $S$-matrices are obtained by LSZ reduction, which leads
to the exact results of the single-photon transmission and the second order
correlation functions of the two out-going photons.

\subsection{Single-photon transport}

For the single-photon case, using the $T$-matrix element (\ref{T}), the
Green function (\ref{G}), and the generating functional (\ref{Z}), we obtain%
\begin{equation}
S_{p;k}=T_{k}\delta _{p,k}+R_{k}\delta _{p,-k},
\end{equation}%
where%
\begin{equation}
T_{k}=1-i\left\vert V_{R}\right\vert ^{2}\left\langle 0\right\vert a\frac{1}{%
k-H_{\mathrm{eff}}}a^{\dagger }\left\vert 0\right\rangle  \label{t1}
\end{equation}%
and%
\begin{equation}
R_{k}=-iV_{L}V_{R}^{\ast }\left\langle 0\right\vert b\frac{1}{k-H_{\mathrm{%
eff}}}a^{\dagger }\left\vert 0\right\rangle  \label{r1}
\end{equation}%
are the transmission and reflection coefficients, respectively. Here,%
\begin{eqnarray}
H_{\mathrm{eff}} &=&\Omega \left\vert e\right\rangle \left\langle
e\right\vert +(\omega _{c}-i\frac{\Gamma _{R}}{2})a^{\dagger }a+(\omega
_{c}-i\frac{\Gamma _{L}}{2})b^{\dagger }b  \notag \\
&&+[hb^{\dagger }a+\sigma ^{+}(g_{a}a+g_{b}b)+\mathrm{H.c.}]\,,
\end{eqnarray}%
is the effective Hamiltonian with $\Gamma _{R}=\left\vert V_{R}\right\vert
^{2}$ and $\Gamma _{L}=\left\vert V_{L}\right\vert ^{2}$ being the decay
rates of the two whispering-gallery modes to the waveguide, respectively.
Without loss of generality, we consider $\Gamma _{R}=\Gamma _{L}\equiv
\Gamma $. We notice that the excitation number $N=\left\vert e\right\rangle
\left\langle e\right\vert +a^{\dagger }a+b^{\dagger }b$ commutes with $H_{%
\mathrm{eff}}$. Thus, for the single-photon scattering, in the
single-excitation subspace as spanned by the basis $\{\left\vert
e\right\rangle \left\vert 0\right\rangle _{a}\left\vert 0\right\rangle
_{b},\left\vert g\right\rangle \left\vert 1\right\rangle _{a}\left\vert
0\right\rangle _{b},\left\vert g\right\rangle \left\vert 0\right\rangle
_{a}\left\vert 1\right\rangle _{b}\}$, we represent the effective
Hamiltonian as%
\begin{equation}
H_{\mathrm{eff}}^{(1)}=\left(
\begin{array}{ccc}
\Omega & g_{a} & g_{b} \\
g_{a}^{\ast } & \omega _{c}-i\frac{\Gamma }{2} & h^{\ast } \\
g_{b}^{\ast } & h & \omega _{c}-i\frac{\Gamma }{2}%
\end{array}%
\right) .  \label{eff1}
\end{equation}

Using the form of $H_{\mathrm{eff}}^{(1)}$ in Eq. (\ref{eff1}), the
single-photon reflection and transmission coefficients can be determined as%
\begin{equation}
R_{k}=\frac{-iV_{L}V_{R}^{\ast }}{D(k)}[g_{a}g_{b}^{\ast }+h(k-\Omega )],
\label{R}
\end{equation}%
and%
\begin{equation}
T_{k}=1-\frac{i\Gamma }{D(k)}[(k-\Omega )(k-\omega _{c}+i\frac{\Gamma }{2}%
)-\left\vert g_{b}\right\vert ^{2}],  \label{Tk}
\end{equation}%
respectively. Here, we define%
\begin{eqnarray}
D(k) &=&(k-\omega _{c}+i\frac{\Gamma }{2})[(k-\Omega )(k-\omega _{c}+i\frac{%
\Gamma }{2})-G_{+}^{2}]  \notag \\
&&-g_{a}^{\ast }g_{b}h-g_{b}^{\ast }g_{a}h^{\ast }-\left\vert h\right\vert
^{2}(k-\Omega ),
\end{eqnarray}%
and%
\begin{equation}
G_{+}=\sqrt{\left\vert g_{a}\right\vert ^{2}+\left\vert g_{b}\right\vert ^{2}%
}.
\end{equation}%
The results (\ref{R}) and (\ref{Tk}) accord with that obtained in Ref. \cite%
{Theo2,Theo3}.

Equations (\ref{R}) and (\ref{Tk}) are applicable in the absence of
intrinsic atomic or cavity dissipation. In the presence of intrinsic
dissipations, the reflection and transmission probabilities $\left\vert
R_{k}\right\vert ^{2}$ and $\left\vert T_{k}\right\vert ^{2}$ are obtained
by substitution $\Omega -i\gamma _{\mathrm{a}}$ and $\omega _{c}-i\gamma _{%
\mathrm{c}}$ for $\Omega $ and $\omega _{c}$ in Eqs. (\ref{R}) and (\ref{Tk}%
), where $\gamma _{\mathrm{a}}$ and $\gamma _{\mathrm{c}}$ are the intrinsic
decay rates of the cavity and the atom.

\subsection{Two-photon transport}

In the subsection, we derive the exact formula for the scattering
wavefunctions and the second order correlation functions of two out-going
photons by LSZ reduction approach. By evaluating Eqs. (\ref{T}), (\ref{G})
and (\ref{Z}), we obtain the $S$-matrix elements in the two-photon Hilbert
space:%
\begin{eqnarray}
S_{p_{1}p_{2};k_{1}k_{2}}^{(\mathrm{R})} &=&R_{k_{1}}R_{k_{2}}(\delta
_{p_{1},-k_{1}}\delta _{p_{2},-k_{2}}+\delta _{p_{1},-k_{2}}\delta
_{p_{2},-k_{1}})  \notag \\
&&-i\frac{V_{L}^{2}V_{R}^{\ast 2}}{2\pi }\delta
_{p_{1}+p_{2},-k_{1}-k_{2}}U_{p_{1}p_{2};k_{1}k_{2}},  \label{R2}
\end{eqnarray}%
and%
\begin{eqnarray}
S_{p_{1}p_{2};k_{1}k_{2}}^{(\mathrm{T})} &=&T_{k_{1}}T_{k_{2}}(\delta
_{p_{1},k_{1}}\delta _{p_{2},k_{2}}+\delta _{p_{1},k_{2}}\delta
_{p_{2},k_{1}})  \notag \\
&&-i\frac{\left\vert V_{R}\right\vert ^{4}}{2\pi }\delta
_{p_{1}+p_{2},k_{1}+k_{2}}W_{p_{1}p_{2};k_{1}k_{2}},  \label{T2}
\end{eqnarray}%
for two reflected photons and transmitted photons, where $\mathrm{R}$ and $%
\mathrm{T}$ refers to transmission and reflection, respectively. In Eqs. (%
\ref{R2}) and (\ref{T2}),%
\begin{eqnarray}
U_{p_{1}p_{2};k_{1}k_{2}}
&=&F_{1}(p_{1},p_{2};k_{1},k_{2})+F_{1}(p_{1},p_{2};k_{2},k_{1})  \label{U}
\\
&&+F_{1}(p_{2},p_{1};k_{1},k_{2})+F_{1}(p_{2},p_{1};k_{2},k_{1}),  \notag
\end{eqnarray}%
and%
\begin{eqnarray}
W_{p_{1}p_{2};k_{1}k_{2}}
&=&F_{2}(p_{1},p_{2};k_{1},k_{2})+F_{2}(p_{1},p_{2};k_{2},k_{1})  \label{W}
\\
&&+F_{2}(p_{2},p_{1};k_{1},k_{2})+F_{2}(p_{2},p_{1};k_{2},k_{1}).  \notag
\end{eqnarray}%
Here, we define%
\begin{eqnarray}
&&F_{1}(p_{1},p_{2};k_{1},k_{2})  \label{F1} \\
&=&\left\langle 0\right\vert b\frac{1}{-p_{1}-H_{\mathrm{eff}}^{(1)}}b\frac{1%
}{k_{1}+k_{2}-H_{\mathrm{eff}}^{(2)}}a^{\dagger }\frac{1}{k_{2}-H_{\mathrm{%
eff}}^{(1)}}a^{\dagger }\left\vert 0\right\rangle  \notag \\
&&+\frac{1}{-p_{1}-k_{1}}\left\langle 0\right\vert b\frac{1}{-p_{1}-H_{%
\mathrm{eff}}^{(1)}}a^{\dagger }b\frac{1}{k_{2}-H_{\mathrm{eff}}^{(1)}}%
a^{\dagger }\left\vert 0\right\rangle ,  \notag
\end{eqnarray}%
and%
\begin{eqnarray}
&&F_{2}(p_{1},p_{2};k_{1},k_{2})  \label{F2} \\
&=&\left\langle 0\right\vert a\frac{1}{p_{1}-H_{\mathrm{eff}}^{(1)}}a\frac{1%
}{k_{1}+k_{2}-H_{\mathrm{eff}}^{(2)}}a^{\dagger }\frac{1}{k_{2}-H_{\mathrm{%
eff}}^{(1)}}a^{\dagger }\left\vert 0\right\rangle  \notag \\
&&+\frac{1}{p_{1}-k_{1}}\left\langle 0\right\vert a\frac{1}{p_{1}-H_{\mathrm{%
eff}}^{(1)}}a^{\dagger }a\frac{1}{k_{2}-H_{\mathrm{eff}}^{(1)}}a^{\dagger
}\left\vert 0\right\rangle ,  \notag
\end{eqnarray}%
where $H_{\mathrm{eff}}^{(1)}$ is defined in Eq. (\ref{eff1}). In the
two-excitation subspace as spanned by the basis \{$\left\vert g\right\rangle
\left\vert 2\right\rangle _{a}\left\vert 0\right\rangle _{b},$ $\left\vert
e\right\rangle \left\vert 1\right\rangle _{a}\left\vert 0\right\rangle _{b},$
$\left\vert g\right\rangle \left\vert 1\right\rangle _{a}\left\vert
1\right\rangle _{b},$ $\left\vert e\right\rangle \left\vert 0\right\rangle
_{a}\left\vert 1\right\rangle _{b},$ $\left\vert g\right\rangle \left\vert
0\right\rangle _{a}\left\vert 2\right\rangle _{b}$\}, we represent the
effective Hamiltonian%
\begin{equation}
H_{\mathrm{eff}}^{(2)}=\left(
\begin{array}{ccccc}
2\alpha & \sqrt{2}g_{a}^{\ast } & \sqrt{2}h^{\ast } & 0 & 0 \\
\sqrt{2}g_{a} & \alpha +\Omega & g_{b} & h^{\ast } & 0 \\
\sqrt{2}h & g_{b}^{\ast } & 2\alpha & g_{a}^{\ast } & \sqrt{2}h^{\ast } \\
0 & h & g_{a} & \alpha +\Omega & \sqrt{2}g_{b} \\
0 & 0 & \sqrt{2}h & \sqrt{2}g_{b}^{\ast } & 2\alpha%
\end{array}%
\right) ,
\end{equation}%
where $\alpha =\omega _{c}-i\Gamma /2$. In general, direct evaluation of $%
U_{p_{1}p_{2};k_{1}k_{2}}$ and $W_{p_{1}p_{2};k_{1}k_{2}}$ can be rather
complicated. However, there are cases where $U_{p_{1}p_{2};k_{1}k_{2}}$ and $%
W_{p_{1}p_{2};k_{1}k_{2}}$ can be obtained exactly in a compact form. We
will discuss such examples in the next section.

The Fourier transformations of matrix elements (\ref{R2}) and (\ref{T2})
result in wavefunctions of the two reflected or transmitted photons:%
\begin{eqnarray}
\psi _{\mathrm{R}}(x_{1},x_{2}) &=&\frac{1}{4\pi }\int
dp_{1}dp_{2}R_{p_{1}p_{2};k_{1}k_{2}}e^{ip_{1}x_{1}+ip_{2}x_{2}}  \notag \\
&=&\frac{1}{2\pi }e^{-iEx_{c}}[R_{k_{1}}R_{k_{2}}\cos (\Delta _{k}x)  \notag
\\
&&+\frac{1}{2}V_{L}^{2}V_{R}^{\ast 2}\int \frac{d\Delta _{p}}{2\pi i}%
e^{i\Delta _{p}x}U_{p_{1}p_{2};k_{1}k_{2}}],  \label{27}
\end{eqnarray}%
and%
\begin{eqnarray}
\psi _{\mathrm{T}}(x_{1},x_{2}) &=&\frac{1}{4\pi }\int
dp_{1}dp_{2}T_{p_{1}p_{2};k_{1}k_{2}}e^{ip_{1}x_{1}+ip_{2}x_{2}}  \notag \\
&=&\frac{1}{2\pi }e^{iEx_{c}}[T_{k_{1}}T_{k_{2}}\cos (\Delta _{k}x)  \notag
\\
&&+\frac{1}{2}\left\vert V_{R}\right\vert ^{4}\int \frac{d\Delta _{p}}{2\pi i%
}e^{i\Delta _{p}x}W_{p_{1}p_{2};k_{1}k_{2}}],  \label{28}
\end{eqnarray}%
respectively. Here, we define the total momentum $E=k_{1}+k_{2}=p_{1}+p_{2}$%
, the relative momenta $\Delta _{k}=(k_{1}-k_{2})/2$ and $\Delta
_{p}=(p_{1}-p_{2})/2$, as well as the center of mass coordinate $%
x_{c}=(x_{1}+x_{2})/2$ and the relative coordinate $x=x_{1}-x_{2}$. Because $%
U_{p_{1}p_{2};k_{1}k_{2}}$ and $W_{p_{1}p_{2};k_{1}k_{2}}$ are both even
functions of $\Delta _{p}$, $\psi _{\mathrm{R}}(x_{1},x_{2})$ and $\psi _{%
\mathrm{T}}(x_{1},x_{2})$ are invariant under the permutation $%
x_{1}\longleftrightarrow x_{2}$, as required since photons are bosons. From
now on, we only focus on the wavefunctions for $x>0$.

The integral in Eqs. (\ref{27}) and (\ref{28}) can be evaluated by analyzing
the analytic properties of the matrix elements $U_{p_{1}p_{2};k_{1}k_{2}}$
and $W_{p_{1}p_{2};k_{1}k_{2}}$. These matrix elements exhibit three poles $%
p_{l}=E/2-\alpha _{l}$ on the upper half plane, where $\alpha _{l}$ is the
eigenvalue of $H_{\mathrm{eff}}^{(1)}$ with $l=1,2,3$. Here, notice that the
eigenvalues of $H_{\mathrm{eff}}^{(2)}$ do not contribute to the poles in $%
U_{p_{1}p_{2};k_{1}k_{2}}$ and $W_{p_{1}p_{2};k_{1}k_{2}}$, because they
only associate with the total energy $E$, as shown in Eqs. (\ref{F1}) and (%
\ref{F2}), and we consider here an incident two-photon state with a fixed
total energy. The residue theorem then leads to the wavefunctions%
\begin{eqnarray}
\psi _{\mathrm{R}}(x_{1},x_{2}) &=&\frac{1}{2\pi }%
e^{-iEx_{c}}[R_{k_{1}}R_{k_{2}}\cos (\Delta _{k}x)  \label{ref} \\
&&+\frac{1}{2}V_{L}^{2}V_{R}^{\ast 2}\sum_{l}\text{Res}%
_{p_{l}}U_{p_{1}p_{2};k_{1}k_{2}}e^{ip_{l}x}],  \notag
\end{eqnarray}%
and%
\begin{eqnarray}
\psi _{\mathrm{T}}(x_{1},x_{2}) &=&\frac{1}{2\pi }%
e^{iEx_{c}}[T_{k_{1}}T_{k_{2}}\cos (\Delta _{k}x)  \label{tr} \\
&&+\frac{1}{2}\left\vert V_{R}\right\vert ^{4}\sum_{l}\text{Res}%
_{p_{l}}W_{p_{1}p_{2};k_{1}k_{2}}e^{ip_{l}x}],  \notag
\end{eqnarray}%
where Res$_{p_{l}}\Phi (\Delta _{p})$ denotes the residue of the function $%
\Phi (\Delta _{p})$ at $p_{l}$.

From the two-photon wavefunctions, the second-order correlations functions
for the reflected and transmitted photons can be obtained as:%
\begin{eqnarray}
g_{\mathrm{R}}^{(2)}(\tau ) &=&\frac{\left\langle \psi _{\mathrm{R}%
}\right\vert l^{\dagger }(x)l^{\dagger }(x+\tau )l(x+\tau )l(x)\left\vert
\psi _{\mathrm{R}}\right\rangle }{\left\vert \left\langle \psi _{\mathrm{R}%
}\right\vert l^{\dagger }(x)l(x)\left\vert \psi _{\mathrm{R}}\right\rangle
\right\vert ^{2}},  \notag \\
g_{\mathrm{T}}^{(2)}(\tau ) &=&\frac{\left\langle \psi _{\mathrm{T}%
}\right\vert r^{\dagger }(x)r^{\dagger }(x+\tau )r(x+\tau )r(x)\left\vert
\psi _{\mathrm{T}}\right\rangle }{\left\vert \left\langle \psi _{\mathrm{T}%
}\right\vert r^{\dagger }(x)r(x)\left\vert \psi _{\mathrm{T}}\right\rangle
\right\vert ^{2}},  \label{g2}
\end{eqnarray}%
where $r(x)$ ($l(x)$) is the Fourier transformation of $r_{k}$ ($l_{k}$).
Equation (\ref{g2}) can be simplified to yield $g_{s}^{(2)}(\tau
)=\left\vert \psi _{s}(x+\tau ,x)\right\vert ^{2}/\int dy\left\vert \psi
_{s}(x,y)\right\vert ^{2}$, where $s$ denote \textquotedblleft $\mathrm{%
R\textquotedblright }$ and \textquotedblleft $\mathrm{T\textquotedblright }$%
. Therefore, our theoretical results on the two-photon wavefunctions can be
compared to experimental measurement of second-order correlation functions
for this system.

In the presence of intrinsic atomic or cavity dissipation, Eqs. (\ref{ref})
and (\ref{tr}) remain valid, provided that we substitute $\Omega -i\gamma _{%
\mathrm{a}}$ and $\omega _{c}-i\gamma _{\mathrm{c}}$ for $\Omega $ and $%
\omega _{c}$ in $U_{p_{1}p_{2};k_{1}k_{2}}$ and $W_{p_{1}p_{2};k_{1}k_{2}}$.

\subsection{Spectrum of the whispering-gallery-atom system}

For this system, its transport properties are closely related to the
spectrum of the resonator coupling to the atom. Therefore, in this section
we provide a brief discussion of the spectrum of the resonator-atom
Hamiltonian $H_{\mathrm{wg}}$ as defined in Eq. (\ref{2}). We consider two
separate cases:

Case (a): As seen in Eq. (\ref{2}), the whispering-gallery resonator by
itself supports a clockwise and a counter clockwise rotating mode, both
coupled to the atom. However, assuming that either%
\begin{equation}
h=0,  \label{Con1}
\end{equation}%
or%
\begin{equation}
h\neq 0,\text{ and }\frac{g_{a}}{g_{b}}=\pm e^{-i\theta _{h}},  \label{Con2}
\end{equation}%
it is then possible form a linear superposition of these modes that is
decoupled from the atom. In Eq. (\ref{Con2}), $\theta _{h}=\arg h$ is the
phase of $h$. In either case, the resonator-atom Hamiltonian can be
rewritten as:%
\begin{equation}
H_{\mathrm{eff}}=\Omega \left\vert e\right\rangle \left\langle e\right\vert
+\omega _{A}A^{\dagger }A+\omega _{B}B^{\dagger }B+(G_{+}\sigma ^{+}A+%
\mathrm{H.c.}),  \label{h}
\end{equation}%
where the operators $A=(g_{a}a+g_{b}b)/G_{+}$ and $B=(g_{b}^{\ast
}a-g_{a}^{\ast }b)/G_{+}$. For the case described by Eq. (\ref{Con1}), we
have $\omega _{A}=\omega _{B}=\omega _{c}-i\Gamma /2$. For the case
described by Eq. (\ref{Con2}), we have $\omega _{A}=\omega _{c}-i\Gamma
/2\pm \left\vert h\right\vert $ and $\omega _{B}=\omega _{c}-i\Gamma /2\mp
\left\vert h\right\vert $.

\begin{figure}[tbp]
\includegraphics[bb=14 366 569 797, width=8 cm, clip]{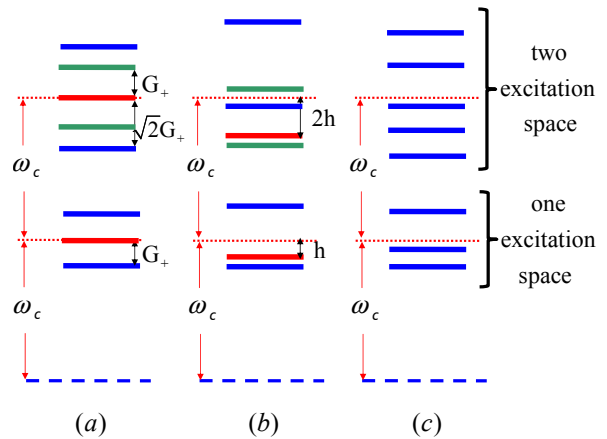}
\caption{(Color online) The schematics for the energy spectra of the
whispering-gallery-atom system: (a) $h=0$. (b) $h=2e^{i\protect\theta_h}$,
and $g_b/g_a=e^{-i\protect\theta_h}$ (c) $h=5i$ and $g_b/g_a=1$. (a) and (b)
corresponds to the effective single mode case, (c) corresponds to the two
mode case. In all panels, dashed blue lines represent the ground state. Red
lines represent a free mode without atom excitation. Notice that it forms a
linear spectrum. Green lines represent mode consists of one excitation in
the free mode, and one excitation in the JC mode. Solid blue lines represent
all other modes.}
\label{fig3}
\end{figure}

It follows from Eq. (\ref{h}) that the mode $B$ is decoupled form the atom,
while mode $A$ couples to the atom through the standard Jaynes-Cummings (JC)
Hamiltonian%
\begin{equation}
H_{\mathrm{JC}}=\Omega \left\vert e\right\rangle \left\langle e\right\vert
+\omega _{A}A^{\dagger }A+(G_{+}\sigma ^{+}A+\mathrm{H.c.}).
\end{equation}%
For subsequent discussion, we refer to mode $A$ as the \textquotedblleft JC
mode\textquotedblright\ and mode B as the \textquotedblleft free
mode\textquotedblright , respectively.

The spectrum of the Hamiltonian (\ref{h}) is displayed in Fig. \ref{fig3}a
for the case $h=0$, and in Fig. \ref{fig3}b for the case when $h\neq 0,$ and
$g_{b}/g_{a}=\pm e^{-i\theta _{h}}$. In either case, due to the very
different nature of modes $A$ and $B$, for the incident photons with energy
on resonance with modes $A$ and $B$, the out-going photons must exhibit very
different statistics behaviors, which could be investigated by the second
order correlation functions $g_{s}^{(2)}(\tau )$.

Case (b): When neither conditions in Eq. (\ref{Con1}) nor Eq. (\ref{Con2})
are satisfied, we can no longer form a linear superposition of the two
whispering gallery modes that decouples with the atom. The energy spectrum,
shown in Fig. \ref{fig3}c, is more complicated.

In the next two sections, we study the photon transmissions for the cases
(a) and (b), respectively. In case (a), the atom effectively couples only to
one of the two modes of the resonator, below we refer to this case as an
\textquotedblleft effective single-mode case\textquotedblright . In
contrast, for case (b) the atom couples to both modes, and below we refer to
this case as a \textquotedblleft two-mode case\textquotedblright .

\section{Results for the effective single-mode case}

In this section, we present transport properties of single-photon and
two-photons, for the effective single-mode case as discussed in Section
III.C, where the resonator supports a photon mode that is decoupled from the
atom. For the numerical results in this and next section, we normalize all
quantities that have the dimension of energy with respect to $\Gamma $, the
waveguide-resonator coupling rate. Also, we choose $\omega _{c}$ as the
origin of the energy axis.

\subsection{Single-photon transport}

\begin{figure}[tbp]
\includegraphics[bb=17 486 575 749, width=8 cm, clip]{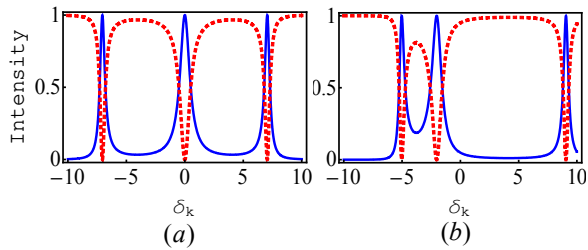}
\caption{(Color online) Single photon transmission (Red dashed curve) and
reflection (Blue solid curve) spectra for the effective single-mode case.
All quantities with energy dimensions are normalized to $\Gamma$, $\protect%
\delta_k$ is the detuning between the incident photon and $\protect\omega_c$%
: (a) $h=0$, $\left\vert g_a\right\vert=\left\vert g_b\right\vert=5$, and $%
\Omega =\protect\omega _{c}$; (b) $\left\vert h\right\vert=2$, $\left\vert
g_a\right\vert=\left\vert g_b\right\vert=5$, $g_b/g_a=e^{-i\protect\theta_h}$%
, and the detuning $\Omega-\protect\omega _{c}=2$ between atom and the
cavity.}
\label{fig4}
\end{figure}

For this effective single-mode case, the single-photon reflection
coefficient (\ref{r1}) can be rewritten as%
\begin{equation}
R_{k}=-i\frac{\Gamma g_{a}g_{b}^{\ast }}{G_{+}^{2}}[\frac{k-\Omega }{%
(k-\Omega )(k-\omega _{A})-G_{+}^{2}}-\frac{1}{k-\omega _{B}}],
\end{equation}%
in which the first and second terms describe the contributions from the JC
mode and the free mode, respectively. We show the single-photon reflection
and transmission probabilities $\left\vert R_{k}\right\vert ^{2}$ and $%
\left\vert T_{k}\right\vert ^{2}$ in Fig. \ref{fig4}. In Fig. \ref{fig4}a,
we choose $h=0$, $\Omega =\omega _{c}$, and $\left\vert g_{a}\right\vert
=\left\vert g_{b}\right\vert $. We find that in the strong coupling limit $%
\Gamma <<G_{+}$ there are three peaks in the reflection spectrum. The peak
in the center of the spectrum corresponds to the free mode $B$. Two other
peaks corresponds to the JC modes in the single-excitation subspace.

In Fig. \ref{fig4}b, we choose $\Omega =\omega _{c}+\left\vert h\right\vert $
and $g_{b}/g_{a}=e^{-i\theta _{h}}$. In the strong coupling limit $\Gamma
\ll G_{+}$, the reflection spectrum again exhibits three peaks. The central
peak again corresponds to the free mode. Compared to the free mode in Fig. %
\ref{fig4}a at the frequency $\omega _{c}$, here the frequency of the free
mode is shifted to $\omega _{c}-\left\vert h\right\vert $ due to the
intermodal coupling. Other two peaks corresponds to the two JC modes.

Based on these results, we conclude that the single-photon transport
consists of two independent scattering processes, i.e., the scattering by
the JC modes, which have atomic excitation, and by the free mode that is
decoupled from the atom.

\subsection{Two-photon transport and photon blockade}

For this effective single-mode case, straightforward calculations of Eqs. (%
\ref{F1}) and (\ref{F2}) lead to the analytic results%
\begin{eqnarray}
U_{p_{1}p_{2};k_{1}k_{2}} &=&-\frac{2g_{b}^{\ast 2}g_{a}^{2}(E-\omega
_{A}-\Omega )}{\prod_{s=\pm }(E-\lambda _{2s})}  \notag \\
&&\frac{(E-2\Omega )(E-2\omega _{A})-4G_{+}^{2}}{\prod_{s=\pm
}\prod_{i=1,2}(k_{i}-\lambda _{1s})(p_{i}+\lambda _{1s})},  \label{u}
\end{eqnarray}%
and%
\begin{eqnarray}
W_{p_{1}p_{2};k_{1}k_{2}} &=&-\frac{2\left\vert g_{a}\right\vert
^{4}(E-\omega _{A}-\Omega )}{\prod_{s=\pm }(E-\lambda _{2s})}  \notag \\
&&\frac{(E-2\Omega )(E-2\omega _{A})-4G_{+}^{2}}{\prod_{s=\pm
}\prod_{i=1,2}(k_{i}-\lambda _{1s})(p_{i}-\lambda _{1s})}.  \label{w}
\end{eqnarray}%
Here, $\lambda _{1\pm }$ and $\lambda _{2\pm }$ are the eigenvalues of $H_{%
\mathrm{JC}}$ in the single- and two-excitation subspaces \cite{MP6}. We
note that $U_{p_{1}p_{2};k_{1}k_{2}}$ and $W_{p_{1}p_{2};k_{1}k_{2}}$, i.e.,
the two-photon correlated scattering, have contributions only from the JC
modes. The free mode does not contribute to the correlated scattering since
it does not have any atomic excitation.

\begin{figure}[tbp]
\includegraphics[bb=20 182 581 749, width=8 cm, clip]{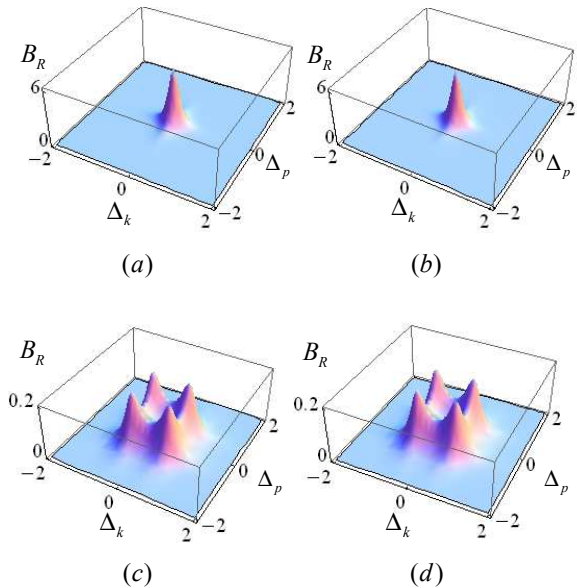}
\caption{(Color online) The two-photon background fluorescence for the
effective single-mode case. All quantities with dimension of energy are
normalized with respect to $\Gamma$: (a) The system parameters are the same
as those in Fig. \protect\ref{fig4}(a), and the detuning $E-2\protect\omega%
_c=-14$; (b) The system parameters are the same as those in Fig. \protect\ref%
{fig4}(b), and the detuning $E-2\protect\omega_c=-10$; (c) The system
parameters are the same as those in Fig. \protect\ref{fig4}(a), and the
detuning $E-2\protect\omega_c=13$; (d) The system parameters are the same as
those in Fig. \protect\ref{fig4}(b), and the detuning $E-2\protect\omega%
_c=17 $.}
\label{fig5}
\end{figure}

In Fig. \ref{fig5}, the two-photon background fluorescence $B_{R}=\left\vert
V_{R}\right\vert ^{4}\left\vert V_{L}\right\vert ^{4}\left\vert
U_{p_{1}p_{2};k_{1}k_{2}}\right\vert ^{2}/4\pi ^{2}$ of the reflected
photons are shown for the strong coupling limit $G_{+}\gg \Gamma $. The
two-photon background fluorescence displays a single peak at $\Delta
_{k}=\Delta _{p}=0$ when the total energy of the incident photons approaches
$2$Re$\lambda _{1\pm }$ (Fig. 4a and b), while the two-photon background
fluorescence splits into four peaks when the total energy of the incident
photons deviates from $2$Re$\lambda _{1\pm }$ (Fig. 4c and d).

In general, the background fluorescence peaks when one of the incident or
outgoing photons has energy that coincides with a single-excitation
eigenstate \cite{MP6}. Examining Eqs. (\ref{u}) and (\ref{w}), we see that
the poles occur at $\Delta _{k}=\pm (E/2-$Re$\lambda _{1s})$ and $\Delta
_{p}=\pm (E/2-$Re$\lambda _{1s})$. Thus, one might expect eight peaks in the
background fluorescence spectra in the general case. However, four of these
poles turn out to have very small residues, resulting in the presence of
only four peaks in Figs. 4c and d.

Together with Eqs. (\ref{u}) and (\ref{w}), it follows from Eqs. (\ref{ref})
and (\ref{tr}) that the outgoing two-photon wavefunctions are%
\begin{equation}
\psi _{\mathrm{R}}(x_{1},x_{2})=\frac{e^{-iEx_{c}}}{2\pi }%
[R_{k_{1}}R_{k_{2}}\cos (\Delta _{k}x)-V_{L}^{2}V_{R}^{\ast 2}g_{b}^{\ast
2}g_{a}^{2}F(x)],  \label{ra}
\end{equation}%
and%
\begin{equation}
\psi _{\mathrm{T}}(x_{1},x_{2})=\frac{e^{iEx_{c}}}{2\pi }[T_{k_{1}}T_{k_{2}}%
\cos (\Delta _{k}x)-\left\vert V_{R}\right\vert ^{4}\left\vert
g_{a}\right\vert ^{4}F(x)],  \label{ta}
\end{equation}%
of the reflected and transmitted photons, respectively, where we define%
\begin{equation}
F(x)=\frac{\sum_{s}s(E-2\lambda _{1s})e^{i(\frac{E}{2}-\lambda
_{1-s})\left\vert x\right\vert }}{\prod_{s=\pm }[(E-\lambda
_{2s})\prod_{i=1,2}(k_{i}-\lambda _{1s})](\lambda _{1+}-\lambda _{1-})}.
\end{equation}

\begin{figure}[tbp]
\includegraphics[bb=19 163 581 656, width=8 cm, clip]{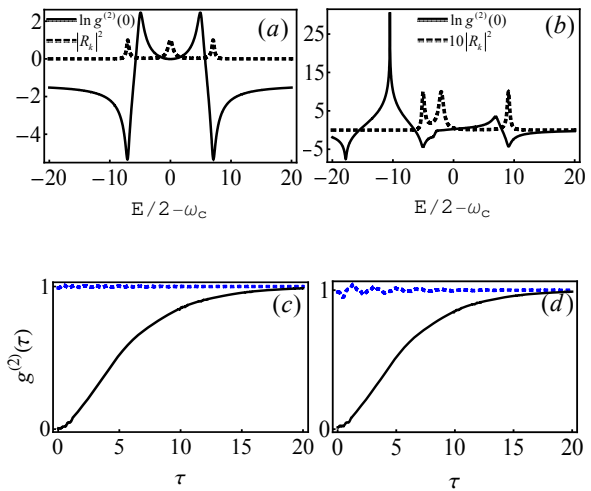}
\caption{(Color online) The second order correlation functions of two
reflected photons for the effective single-mode case. The two incident
photons have the same energy $E/2$. All quantities with dimension of energy
are normalized with respect to $\Gamma$, and the system parameters for the
left and the right panels are the same as those in Fig. \protect\ref{fig4}%
(a) and (b), respectively. In (a) and (b), the second order correlation
function and the reflection coefficient are depicted by the solid and dashed
curves, respectively. (c) The solid (black) and dashed (blue) curves denote $%
g_{\mathrm{R}}^{(2)}(\protect\tau )$ for $E/2-\protect\omega_c=\pm 7$
(single photon being resonant with the JC mode) and $E/2-\protect\omega_c=0$
(single photon being resonant with the JC mode), respectively. (d) The solid
(black) and blue (dashed) curves denote $g_{\mathrm{R}}^{(2)}(\protect\tau )$
for $E/2-\protect\omega_c=-5,9$ (single photon being resonant with the JC
mode) and $E/2-\protect\omega_c=-2$ (single photon being resonant with the
JC mode), respectively.}
\label{fig6}
\end{figure}

In Fig. \ref{fig6}, we plot the second-order correlation for the two
reflected photons, as obtained by applying Eq. (\ref{g2}) to the
wavefunctions determined from Eqs. (\ref{ra}) and (\ref{ta}). Here, we
consider the same two systems in Fig. \ref{fig4}, and assume that the two
incident photons have the same single-photon energy of $E/2$. In Fig. \ref%
{fig4}, we saw that the systems exhibit strong resonant reflection for
single photon, when the single-photon energy is resonant with either the JC
modes or the free modes. However, in the presence of two incident photons,
the statistics of the outgoing photons at these resonances are very
different. As we see in Fig. 5, when the single-photon energy coincides with
the energy of one of JC modes, e.g., $E/2-\omega _{c}\sim -7$ in Fig. \ref%
{fig6}c and $E/2-\omega _{c}\sim -5$ in Fig. \ref{fig6}d, we observe
pronounced anti-bunching behavior with $g_{\mathrm{R}}^{(2)}(0)<g_{\mathrm{R}%
}^{(2)}(\tau )\leq 1$, and therefore a strong photon-blockade effect. On the
other hand, no anti-bunching behavior or photon blockade effect is observed
when the single-photon energy coincides with that of the free mode, e.g., $%
E/2-\omega _{c}\sim 0$ in Fig. \ref{fig6}c and $E/2-\omega _{c}\sim -2$ in
Fig. \ref{fig6}d.

The presence or absence of photon blockade effect at the different
resonances is closely related to energy spectrum as we analyzed in Figs. 2a
and b. Since the free mode $B$ is decoupled from the atom, its energy
spectrum $E=n(\omega _{c}\mp \left\vert h\right\vert )$ is linear. Thus,
there is no photon blockade effect when the incident photon is resonant with
the free mode. On the other hand, the spectrum of the JC mode is highly
non-linear. Thus, while a single-photon on resonance with one of the JC
modes is reflected, two such photons can not be simultaneously reflected
since the total energy is off resonance in the two-excitation subspace.

\section{Results for the two-mode case}

\begin{figure}[tbp]
\includegraphics[bb=26 449 560 736, width=8 cm, clip]{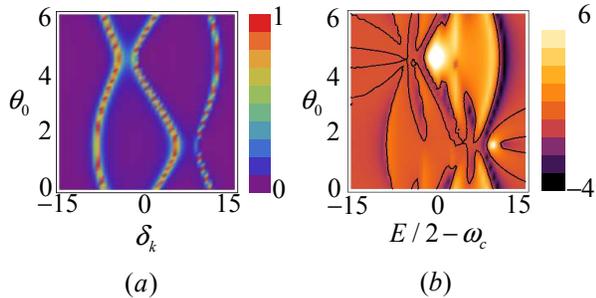}
\caption{(Color online) The single photon reflection and the second order
correlation functions, where $\Gamma$ is taken as units, $\left\vert
h\right\vert=5$, $\protect\theta_{h}=\protect\pi/2$, and other parameters
are the same as those in Fig. \protect\ref{fig4}(b): (a) The single photon
reflection; (b) The second order correlation functions $\ln g^{(2)}_{\mathrm{%
R}}(0)$. Here, the black curves denote $g^{(2)}_{\mathrm{R}}(0)=1.0$.}
\label{fig7}
\end{figure}

In this section, we use the general formula derived in Sec. III to study the
single- and two-photon transports for the two-mode case. In the presence of
the intermodal coupling, i.e., $h\neq 0$, we see in Eq. (\ref{Con2}) that
except for a very special choice of the relative phase of $g_{a}$ and $g_{b}$%
, corresponding to one special choice of atom position, in general one can
not form a photon mode that decouples from the atom. We define $\theta
_{0}=arg(g_{b}/g_{a})$. The analytic results for the single-photon intensity
response function, and the two-photon correlation function, depends only on $%
\theta _{0}+\theta _{h}$. Thus, in the numerical results, without loss of
generality we fix $\theta _{h}$ $=\pi /2$, and vary $\theta _{0}$ from $0$
to $2\pi $.

In the two-mode\ case, where atom couples to two photon modes, both the
single photon reflection (Fig. 6a), and in the two-photon case the
statistics of the outgoing photon (Fig. 6b), becomes dependent upon $\theta
_{0}$, and hence the position of the atom. Therefore, by controlling the
position of the atom, one can tune both the single-photon transport and
two-photon correlation properties.

\begin{figure}[tbp]
\includegraphics[bb=34 539 573 779, width=8 cm, clip]{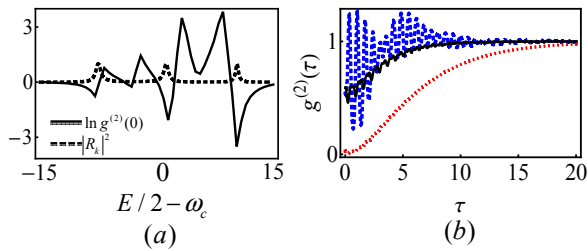}
\caption{(Color online) The second order correlation functions of two
reflective photons, where $\left\vert h\right\vert =5$, $\protect\theta _{h}=%
\protect\pi /2$, $\protect\theta _{0}=0$, and $\Gamma $ is taken as units:
(a) The solid and dashed curves denote $\ln g_{\mathrm{R}}^{(2)}(0)$ and the
single photon reflection; (b) The solid (black), dashed (blue) and dotted
(red) curves denote the $g_{\mathrm{R}}^{(2)}(\protect\tau )$ for the
incident resonant energies $E/2-\protect\omega _{c}=-7.39$, $1$ and $10.84$,
respectively. Other parameters are the same as those in Fig. \protect\ref%
{fig4}(b).}
\label{fig9}
\end{figure}

Similar to the effective one-mode case, here with the choice of parameters
that place the system in the strong-coupling regime, the single-photon
reflection also exhibits three peaks (Fig. 7a, dashed line). Unlike the
effective one-mode case, however, here all three peaks exhibit photon
blockade effect. In Fig. 7b, we plot $g^{(2)}(0)$ for the two reflected
photons, when two photons having the same energy $E/2$ are incident upon the
system. We see strong resonant behavior of $g^{(2)}(0)$ at the energy
corresponding to the three single-photon eigenmodes. The $g^{(2)}(\tau )$ at
these three energies are plotted in Fig. 7c, where we see strong photon
blockade effect with $g^{(2)}(0)\ll 1$. In the two-mode case, all eigenmodes
have atom excitation, and hence contributes to correlated transport.

\section{Final Remark and Conclusion}

As a final remark, in the numerical examples we have focused on the case
where there is no intrinsic dissipation of either the atom or the cavity. In
general, the dissipations suppress the single-photon reflection and the
photon blockade effect. We expect that due to the small decay rates of the
whispering-gallery system to the environment, the photon blockade effect is
well-protected.

In summary, in this paper we study the two-photon transport of the
whispering-gallery-atom system by LSZ reduction approach. We consider the
cases of systems with or without intermodal mixing, and present exact
results on the second-order correlation functions of the two reflected
photons, which exhibit photon-blockade effect. We expect the LSZ formalism
may be developed to treat photon-blockade effect in other systems as well,
including the opto-mechanical system that was considered in \cite%
{om,ZP,XF,Rabl}.

%%%%%%%%%%%%%%%%%%%%%%%%%%%%%%%%%%%%%%%%%%%%%%%%%%%%%%%%%%%%%%%%%%%%%%%
\acknowledgments Tao Shi was supported by the EU project AQUTE. Shanhui Fan acknowledges the support of an AFOSR-MURI program on quantum metamaterial.
%%%%%%%%%%%%%%%%%%%%%%%%%%%%%%%%%%%%%%%%%%%%%%%%%%%%%%%%%%%%%%%%%%%%%%%

\end{document}